\documentstyle[12pt]{article}

\def\hybrid{\topmargin -20pt    \oddsidemargin 0pt  
        \headheight 0pt \headsep 0pt  
        \textwidth 6.25in       % A4 paper  
        \textheight 9.5in       % A4 paper  
        \marginparwidth .875in  
        \parskip 5pt plus 1pt   \jot = 1.5ex}  
\def\noi{\noindent}  
  
\def\baselinestretch{1.2}  
  
\catcode`\@=11  
  
\def\marginnote#1{}  
\def\draftlabel#1{{\@bsphack\if@filesw {\let\thepage\relax  
   \xdef\@gtempa{\write\@auxout{\string  
      \newlabel{#1}{{\@currentlabel}{\thepage}}}}}\@gtempa  
   \if@nobreak \ifvmode\nobreak\fi\fi\fi\@esphack}  
        \gdef\@eqnlabel{#1}}  
\def\@eqnlabel{}  
\def\@vacuum{}  
\def\draftmarginnote#1{\marginpar{\raggedright\scriptsize\tt#1}}  
  
\def\draft{\oddsidemargin -.2truein  
        \def\@oddfoot{\sl preliminary draft \hfil  
        \rm\thepage\hfil\sl\today\quad\militarytime}  
        \let\@evenfoot\@oddfoot \overfullrule 3pt  
        \let\label=\draftlabel  
        \let\marginnote=\draftmarginnote  
   \def\@eqnnum{(\theequation)\rlap{\kern\marginparsep\tt\@eqnlabel}%  
\global\let\@eqnlabel\@vacuum}  }  
  
\def\preprint{\twocolumn\sloppy\flushbottom\parindent 2em  
        \leftmargini 2em\leftmarginv .5em\leftmarginvi .5em  
        \oddsidemargin -.5in    \evensidemargin -.5in  
        \columnsep .4in \footheight 0pt  
        \textwidth 10.in        \topmargin  -.4in  
        \headheight 12pt \topskip .4in  
        \textheight 6.9in \footskip 0pt  
        \def\@oddhead{\thepage\hfil\addtocounter{page}{1}\thepage}  
        \let\@evenhead\@oddhead \def\@oddfoot{} \def\@evenfoot{} }  
  
\def\numberbysection{\@addtoreset{equation}{section}  
        \def\theequation{\thesection.\arabic{equation}}}  
  
\def\underline#1{\relax\ifmmode\@@underline#1\else  
        $\@@underline{\hbox{#1}}$\relax\fi}

\def\titlepage{\@restonecolfalse\if@twocolumn\@restonecoltrue  
\onecolumn  
     \else \newpage \fi \thispagestyle{empty}\c@page\z@  
        \def\thefootnote{\fnsymbol{footnote}} }  
  
\def\endtitlepage{\if@restonecol\twocolumn \else \newpage \fi  
        \def\thefootnote{\arabic{footnote}}  
        \setcounter{footnote}{0}}  %\c@footnote\z@ }  
  
\catcode`@=12  
\relax  
  
\def\figcap{\section*{Figure Captions\markboth  
        {FIGURECAPTIONS}{FIGURECAPTIONS}}\list  
        {Figure \arabic{enumi}:\hfill}{\settowidth\labelwidth{Figure  
999:}  
        \leftmargin\labelwidth  
        \advance\leftmargin\labelsep\usecounter{enumi}}}  
 \relax  
\def\tablecap{\section*{Table Captions\markboth  
        {TABLECAPTIONS}{TABLECAPTIONS}}\list  
        {Table \arabic{enumi}:\hfill}{\settowidth\labelwidth{Table  
999:}  
        \leftmargin\labelwidth  
        \advance\leftmargin\labelsep\usecounter{enumi}}}  
 \relax  
\def\reflist{\section*{References\markboth  
        {REFLIST}{REFLIST}}\list  
        {[\arabic{enumi}]\hfill}{\settowidth\labelwidth{[999]}  
        \leftmargin\labelwidth  
        \advance\leftmargin\labelsep\usecounter{enumi}}}  
 \relax  
\makeatletter  
\newcounter{pubctr}  
\def\publist{\@ifnextchar[{\@publist}{\@@publist}}  
\def\@publist[#1]{\list  
        {[\arabic{pubctr}]\hfill}{\settowidth\labelwidth{[999]}  
        \leftmargin\labelwidth  
        \advance\leftmargin\labelsep  
        \@nmbrlisttrue\def\@listctr{pubctr}  
        \setcounter{pubctr}{#1}\addtocounter{pubctr}{-1}}}  
\def\@@publist{\list  
        {[\arabic{pubctr}]\hfill}{\settowidth\labelwidth{[999]}  
        \leftmargin\labelwidth  
        \advance\leftmargin\labelsep  
        \@nmbrlisttrue\def\@listctr{pubctr}}}  
 \relax  
\makeatother  
\newskip\humongous \humongous=0pt plus 1000pt minus 1000pt

\newif\ifdtup

\font\Scbig=cmss10 scaled\magstep1  
\font\Scscr=cmss8 scaled\magstep1  
\font\Scscrscr=cmss8  
\newfam\Scfam  
\textfont\Scfam=\Scbig  
\scriptfont\Scfam=\Scscr  
\scriptscriptfont\Scfam=\Scscrscr

\relax  
\hybrid  
\def\lvm{\leavevmode\hbox to\parindent{\hfill}}  
  
\def\thefootnote{\fnsymbol{footnote}}  
\def\BE{\begin{equation}}  
\def\EE{\end{equation}}  
\def\BA{\begin{eqnarray}}  
\def\EA{\end{eqnarray}}

\def\tt{\bar\tau}  
  
\def\lvm{\leavevmode\hbox to\parindent{\hfill}}  
\def\bar{\overline}

\def\BE{\begin{equation}}  
\def\EE{\end{equation} \vskip 0.30\baselineskip}  
\def\BA{\begin{array}}  
\def\EA{\end{array}}  
  
\def\noi{\noindent}  
  
\def\frac#1#2{{\textstyle{{#1}\over{#2}}}}

\newif\ifold \oldtrue   
\let\ssection=\section  
\def\section{\setcounter{equation}{0}\ssection}  
  
\begin{document}  
\renewcommand{\theequation}{\arabic{equation}}  
\newcommand{\beq}{\begin{equation}}  
\newcommand{\eeq}[1]{\label{#1}\end{equation}}  
\newcommand{\ber}{\begin{eqnarray}}  
\newcommand{\eer}[1]{\label{#1}\end{eqnarray}}  
\begin{titlepage}  
\begin{center}  
   
\hfill IMAFF-FM-04/20\\
%\hfill gr-qc/0605096
\vskip 1in
  
{\large \bf The Fermi Paradox in the light of the Inflationary and 
Brane World Cosmologies} 
\vskip 1in 
 
{\bf Beatriz Gato-Rivera}\\
\vskip .5in

{\em Instituto de Matem\'aticas y F\'\i sica Fundamental, CSIC \\
Serrano 123, Madrid 28006, Spain}\\

\vskip 1in

\end{center}
  
\begin{center} {\bf ABSTRACT } \end{center}  
\begin{quotation}  
The Fermi Paradox is discussed in the light of the inflationary and brane world 
cosmologies. We conclude that some brane world cosmologies may be of relevance 
for the problem of civilizations spreading across our galaxy, strengthening 
the Fermi Paradox, but not the inflationary cosmologies, as has been proposed.
   
\end{quotation} 
 
\vskip 2cm  
  
December 2004  

\vskip 2cm

\noi 
Published in the volume `Trends in General Relativity and Quantum Cosmology',
Nova Science Eds., New York, 2006.

\end{titlepage}  
  
\def\baselinestretch{1.2}  
\baselineskip 17 pt

\section{The Fermi Paradox}\lvm  
  
 Los Alamos, summer 1950. At lunch Enrico Fermi, Edward Teller and other
colleagues bring up the subject of unidentified flying objects, 
which was very popular at
that time. After a while, when they had changed subjects Fermi suddenly asked:
Where is everybody? Performing fast mental computations, Fermi had reached the
conclusion that alien civilizations should have been around visiting Earth for many
thousands or millions of years. Therefore, why we do not see them? This is the
Fermi Paradox.

Although Fermi never explained how he made his computations, nor gave an
estimate of the number of civilizations which should have visited Earth, he had to
rely on arguments like these: In our galaxy there are thousands of millions of 
stars much older than the Sun, many of them thousands of millions of years older 
(in the `habitable zone' of the galaxy they are on average one thousand 
million years older  \cite{LFG}). 
Therefore many civilizations must have arisen in our galaxy before ours and 
a fraction of them must have expanded through large regions of the galaxy
or even through the entire galaxy.
 
Some other arguments that probably were not available at that time involve 
estimates about the lifetime of the second generation stars, inside of which the 
chemical elements of organic matter are made, and also involve estimates of
the total time necessary for a technological civilization to colonize, or visit, the
whole galaxy. Regarding the second generation stars, they are formed only two
million years after the supermassive first generation stars. The reason is that
supermassive stars burn out completely exploding as supernovae in one 
million years only and  it takes another million years for the debris to form new
stars.  Therefore the appearance of organic matter in our galaxy could have happened 
several thousands of millions of years before the Sun came into existence. 
As to the total time necessary to colonize, or visit, the whole galaxy by a technological
civilization, conservative computations of diffusion modeling give estimates from 
5 to 50 million years \cite{Sci}, which is a cosmologically short timescale. Besides 
these considerations, the fact that life on Earth started very early supports the 
views, held by many scientists, that life should be abundant in the Universe.

\section{Solutions to the Fermi Paradox}\lvm

Many solutions have been proposed to the Fermi paradox. We classify them
as expansionist and non-expansionist, depending on whether they
rely on the idea that technological civilizations expand or do not expand through
large regions of the galaxy. The most popular non-expansionist solutions, based 
on the assumption that technological civilizations do not expand beyond a small 
neighborhood, are the following ones:

\begin{itemize}

\item
Interstellar travel is not possible no matter the scientific and technological
level reached by a civilization. Advocates of this idea are, for example, most
experts of the SETI (Search for Extraterrestrial Intelligence) project, who 
for about 30 years are trying to detect electromagnetic signals from distant
civilizations. 

\item 
Generically, advanced civilizations have little or no interest in expanding 
through large regions of the galaxy.

\item 
Technological civilizations annihilate themselves, or disappear by natural 
catastrophes, before having the chance to spread through large regions of the galaxy.

\end{itemize}

\vskip .2in

On the other hand, the most popular expansionist solutions to the Fermi Paradox,
based on the assumption that generically technological civilizations 
(or a non-negligible fraction of them) do expand through large regions of the 
galaxy, are the following ones:

\begin{itemize}

\item 
Alien civilizations do visit Earth at present times, for different purposes, and/or 
have visited Earth in the past. In this respect it is 
remarkable the fact that Francis Crick, one of the discoverers of the DNA 
structure, proposed in the mid-seventies that life on Earth could have been
inseminated on purpose by alien intelligences intelligences\footnote{It is less 
known that several years before Crick, in 1960, the astronomer Thomas 
Gold suggested, during a congress in Los Angeles, that space travellers 
could have brought life to Earth some thousands of millions of years ago.}. 
Besides, some scientists as well as many authors of popular books, have 
speculated that some unidentified flying objects could be true alien spacecrafts 
whereas some `gods' descending from the sky, abundant in many ancient traditions, 
could have been just alien astronauts (see for example \cite{Deardorff}). 

\item 
Advanced alien civilizations have not encountered the Solar System yet, but
they are on their way.

\item 
Advanced alien civilizations might have strong ethical codes against interfering
with primitive life-forms \cite{Sagan}.

\item 
Advanced aliens ignore us because of lack of interest due to our low primitive 
level. For example Robert Jastrow, ex-director of Mt. Wilson Observatory, 
claims \cite{RJ} that, in average, advanced civilizations should consider us 
as larvae due to the fact that they should be thousands of millions of years
ahead of us.... and who would be interested in communicating with larvae?

\item 
Alien civilizations have not reached us yet because intelligent life is extremely 
difficult to emerge. Otherwise alien civilizations would necessarily be here.
As a result we could find ourselves among the most evolved
technological civilizations in our galaxy or we could even be the only one. 

\end{itemize}

\vskip .2in 

Besides these simple solutions there are many more exotic proposals\footnote{As 
many as fifty solutions to the Fermi Paradox have been collected in the book
\cite{SW}, although there are several left out which we mention in this article.}. 
For example,
a rather drastic expansionist solution is given by the string theoretist Cumrun Vafa
who thinks that the fact that we do not see aliens around could be the first proof 
of the existence of brane worlds: all advanced aliens would have emigrated to 
better parallel universes (our Universe would have zero measure) \cite{CuVa}. 

Recently we made our own proposal for solving the Fermi Paradox \cite{article}.
It states that, at present, all the typical galaxies of the Universe are already
colonized (or large regions of them) by advanced civilizations, a small proportion
of their individuals belonging to primitive subcivilizations, like ours. That is, we put 
forward the possibility that our small terrestrial civilization is embedded in a large 
civilization unknowingly and this situation should be common in all typical galaxies. 
Whether the primitive subcivilizations would know or ignore their 
low status would depend, most likely, on the ethical standards  
of the advanced civilization in which they are immersed. If the standards 
were low, the individuals of the primitive subcivilizations would be 
surely abused in many ways. Consequently, in this case the primitive  
individuals would be painfully aware of their low status. If the 
ethical standards of the advanced individuals were high instead, 
then very probably they would respect the natural evolution (social, cultural) 
of the primitive subcivilizations, treating them
`ecologically' as some kind of protected species. In this case,  
which could well describe the situation of the terrestrial civilization, 
the primitive individuals would be completely unaware of the existence 
of the large advanced civilization.
Observe that the `alien visitors', from the viewpoint of the primitive individuals,
would not be so from the viewpoint of the advanced individuals because 
they rather would be visiting, or working in, their own territory. Observe also
that we do not postulate that advanced alien civilizations might have strong
ethical codes against interfering with primitive civilizations. We simply 
distinguish between aggressive and non-aggresive advanced civilizations,
which in our opinion is a much more realistic idea. In this respect, the fact that our
civilization has never been attacked by aggressive aliens, as far as history
knows, could well be a clue that we belong to a non-aggressive advanced
civilization which protects planet Earth, as part of its territory.

If this scenario were true for our civilization, then the {\it Subanthropic 
Principle} \cite{article} would also hold. It states that
we are not typical among the intelligent observers from the Universe. 
Typical civilizations of typical galaxies would  be hundreds of thousands, 
or millions, of years more evolved than ours and, consequently, typical intelligent 
observers would be orders of magnitude more intelligent than us.  Finally,
in order for our proposal to be a solution of the Fermi Paradox, we complement 
it with an additional hypothesis, called the {\it Undetectability Conjecture} 
\cite{article}, which explains why we do not detect any signals of civilization 
from the outer space. This conjecture states that, generically, all advanced 
enough civilizations camouflage their planets for security reasons, 
because of the existence of aggressive advanced civilizations, so that  
no sign of civilization can be detected by external observers, who  
would only obtain distorted data for disuasion purposes.

\section{The Fermi Paradox in the light of the inflationary and
brane world cosmologies}\lvm
  
\subsection{Inflationary cosmologies}

Almost two years ago Ken D. Olum argued \cite{Olum} that in the infinite Universe 
predicted by eternal inflation there must be some large civilizations which have 
spread widely through the Universe and contain a huge number of individuals.
Although the Fermi Paradox was not mentioned, the underlying idea was
again that in the observable Universe, because of the existence of thousands
of billions of stars much older than the Sun, there must be large civilizations 
much older than ours. Then the author presented some computations regarding
the probabilities that typical intelligent observers belong to a large (galactic
size) civilization at the present time. In particular, using the assumption of 
an infinite Universe, like in the models of eternal inflation, and doing some
conservative computations he predicted that  `{\it all but one individual in $10^8$ 
belongs to a large civilization}'.

Dropping the infinite Universe assumption, 
but keeping still inflation, the author claimed that the predictions are not 
very different than for the previous case because inflationary models, even
if not eternal, usually produce a Universe much larger that the Universe we 
observe. Then, invoking the anthropic premise that we are typical individuals, 
he predicted that there is a probability of $10^8$ versus 1 that we belong to 
a large civilization, in conflict with observation. The author concluded:  
`{\it A straightforward application of anthropic reasoning and reasonable  
assumptions about the capabilities of other civilizations predict that we  
should be part of a large civilization spanning our galaxy. Although the  
precise confidence to put in such a prediction depends on one's  
assumptions, it is clearly very high. Nevertheless, we do not belong  
to such a civilization. Thus something should be amiss....... but then  
what other mistakes are we making.....?}'  According to our proposal we 
could be part of a large civilization spanning our galaxy, or a large region of it, 
without being aware, because of our primitive low status together with the 
high ethics of our hosts, as discussed in the previous section. 
The two major mistakes of the author, therefore, 
would have been to assume: first, that we are typical intelligent observers, 
and, second, that to belong to a civilization implies to be a citizen of it.  

Olum's article seems to make the Fermi Paradox even stronger. However we do 
not agree with these views because, in our opinion, only our own galaxy matters 
for the problem at hand, regarding the observable Universe. Any other 
galaxies are much too far to be even considered as candidates for visitation 
or colonization in either direction (our closest neighbour Andromeda is more 
than two million light years away). As a result, it should not matter whether or not 
there is inflation or whether or not the Universe is infinite or finite, as long as 
inflation and the age of the Universe have little influence on the age of our own
galaxy, in particular on the age of the Sun and the thousands of millions of
stars older than the Sun. In other words, cosmological inflation affects the 
large scale structure of the Universe, but not the small galaxy-size scale, which
is the only relevant scale for the spreading of civilizations. Otherwise one would
have to postulate the existence of exotic phenomena, like wormholes \cite{worm},
Alcubierre warp drive \cite{Warpd},  Krasnikov tubes \cite{Kras}, etc., with the hope
that they would allow intergalactic travel by advanced civilizations endowed with
the appropriate technology. Our honest opinion, however, is that if this type of 
phenomena really existed, they could perhaps offer some practical help regarding 
interstellar travel in our observable Universe, but only {\it inside} our own galaxy.

\subsection{Brane world cosmologies}

In the last years brane world models have been of increasing interest for
both Particle Physics and Cosmology. They put forward the possibility that our
Universe is located in a subspace (brane) of a higher dimensional spacetime,
with the standard model fields confined on the brane and gravity
propagating in the bulk. This allows large, and even infinite, extra dimensions.
Recent work on brane worlds started following the proposals of 
N. Arkani-Hamed, S. Dimopoulos and G.R. Dvali \cite{B-W}, and 
L. Randall and R. Sundrum \cite{RanSun}. The interest in these models
comes from the fact that they offer the possibility to solve, or view 
from a newly different perspective, many longstanding problems in
Particle Physics and Cosmology. First of all brane worlds may solve the
hierarchy problem between the electroweak scale $M_W$ and the Planck
scale $M_{Pl}$ without the need to introduce supersymmetry: $M_W$
would be the only fundamental scale in nature, and the weakness of
the 4-dimensional gravity would be just a consequence of the 4-dimensional
graviton wave function being diluted in the bulk. But brane worlds can also
shed some light on the baryogenesis and leptogenesis, on the proton stability,
on the small masses and large mixing of the neutrinos, on the gravitational
lensing (by brane world black holes), on the nature of the dark matter and
dark energy, etc. (see \cite{reviews} for brane world reviews).  

In contrast with the inflationary cosmologies, brane world cosmologies have 
the potential to truly strengthen the Fermi Paradox. The reason is that in the 
brane world scenarios our observable Universe is embedded in a much larger 
cosmos with, at least, one more large spatial dimension. Along the
large extra spatial dimensions there may be other universes which could
be parallel to our own, or intersecting it somewhere. 
If any of these scenarios turns out to describe the real world, then it would be 
natural to expect that some of these universes would have the 
same laws of Physics as ours and many of the corresponding advanced 
civilizations could master techniques to travel or `jump' through the extra 
dimensions for visitation or colonization purposes. Moreover, one has to take
into account that many of these universes could be very close to ours,
even at less than one millimeter distance along an extra dimension.  
This opens up enormous possibilities 
regarding the expansion of advanced civilizations simultaneously through 
several parallel universes with the same laws of Physics, resulting in 
multidimensional empires. It could even happen that the expansion to 
other `parallel' galaxies through extra dimensions could be easier, 
with lower cost, than the expansion inside one's own galaxy.

In many other universes, however, the laws of Physics would be different,
corresponding perhaps to different vacua of the `would be' ultimate Theory of 
Everything, resulting probably in `shadow matter' universes with respect to ours.  
This means that shadow matter would only interact with our matter 
gravitationally, in the case it would be brought to our Universe  
using appropriate technology. This does not mean, however, that the shadow 
universes would be necessarily empty of intelligent beings. If some of them 
had advanced civilizations, some of their individuals could even `jump' to our 
Universe, but not for colonization purposes since they would not even see
our planets and stars, which they would pass through almost unaware (they
would only notice the gravitational pull towards their centers). And the other 
way around, we could neither see, nor talk to, the shadow visitors, although
they could perhaps try to communicate with the `would be' intelligent beings 
of our Universe through gravitational waves, for example. 

At present we are still in a very premature phase in the study of brane worlds 
and we do not know whether these ideas are in fact realistic. Nevertheless, the
idea of large extra dimensions and parallel universes is acquiring greater and 
greater importance in the scientific community, among both theoreticians and 
experimentalists. As a matter of fact, the experimental signatures 
expected from large extra 
dimensions, at present and future colliders, are well understood by now
\cite{colliders} and an intense experimental search is currently under way.
For example, experiments starting in 2007 at the LHC (CERN) will be looking,
among other events, for signatures of large extra dimensions.

\section{Conclusions and Final Remarks}  
 
We have discussed whether the inflationary and brane world cosmologies
have the potential to influence the problem known as the Fermi Paradox. 
We conclude that cosmological inflation does not influence the spreading 
of civilizations through our galaxy, contrary to recent claims, since it does
not affect the age of the stars in it. Even if inflation produces a much older,
or infinite Universe, like in eternal inflation models, it will be only relevant at
very large scales far beyond visitation or colonization by technological
civilizations, unless one invokes very exotic phenomena able to connect
distant galaxies efficiently enough with regards to space travel (possibility
in which we, honestly, do not believe). Only in this case an older, 
or infinite Universe would increase the probabilities of visitation by
civilizations of other galaxies in our observable Universe. 

In the case of brane world cosmologies we conclude that some of these
scenarios could have the potential to strengthen the Fermi Paradox, provided
they involved parallel universes with the same laws of Physics as ours. 
Then it would be natural to expect the existence of advanced civilizations
capable of traveling through extra dimensions for visitation or colonization
purposes, in either direction. It could even happen that the expansion to
other parallel galaxies through extra dimensions could be easier, with
lower energetic cost, than the expansion inside one's own galaxy, since
many of these universes could be very close to ours,
even at less than one millimeter distance along an extra dimension.

\vskip 1cm   

\noi
{\bf Acknowledgements}

I am indebted to Pedro Gonz\'alez-D\'\i az for several interesting discussions.

\vskip .2in 
\noi

\end{document}